\documentclass[aps,prl,showpacs,floatfix,twocolumn,superscriptaddress]{revtex4-1}

\usepackage[pdftex]{graphicx}
\usepackage{amsmath}
\usepackage{hyperref}
\usepackage{amssymb,bm}
\usepackage{braket}

\begin{document}
\title{Unconditional quantum-noise suppression via measurement-based quantum feedback%
}
\author{Ryotaro~Inoue}
\affiliation{Department of Physics, Graduate School of Science, Kyoto University, Kitashirakawa Oiwake-Cho, Kyoto 606-8502, Japan}
\affiliation{CREST, JST, 4-1-8 Honcho Kawaguchi, Saitama 332-0012, Japan}
\author{Shin-Ichi-Ro~Tanaka}
\affiliation{Department of Physics, Graduate School of Science, Kyoto University, Kitashirakawa Oiwake-Cho, Kyoto 606-8502, Japan}
\author{Ryo~Namiki}
\affiliation{Department of Physics, Graduate School of Science, Kyoto University, Kitashirakawa Oiwake-Cho, Kyoto 606-8502, Japan}
\author{Takahiro~Sagawa}
\affiliation{The Hakubi Center for Advanced Research, Kyoto University, Yoshida-Ushinomiya-cho, Sakyo-ku, Kyoto 606-8302, Japan}
\affiliation{Yukawa Institute of Theoretical Physics, Kyoto University, Kitashirakawa Oiwake-Cho, Kyoto 606-8502, Japan}
\author{Yoshiro~Takahashi}
\affiliation{Department of Physics, Graduate School of Science, Kyoto University, Kitashirakawa Oiwake-Cho, Kyoto 606-8502, Japan}
\affiliation{CREST, JST, 4-1-8 Honcho Kawaguchi, Saitama 332-0012, Japan}

\begin{abstract}

We demonstrate unconditional quantum-noise suppression in a collective spin system via feedback control based on quantum non-demolition measurement (QNDM).
We perform shot-noise limited collective spin measurements on an ensemble of $3.7\times 10^5$ laser-cooled ${}^{171}\text{Yb}$ atoms in their spin-1/2 ground states.
Correlation between two sequential QNDMs indicates  $-0.80^{+0.11}_{-0.12}\,\mathrm{dB}$ quantum noise suppression in a conditional manner.
Our feedback control successfully converts the conditional quantum-noise suppression into the unconditional one without significant loss of the noise reduction level.

\end{abstract}
\maketitle

Feedback is an essential building block of classical control procedure.
One can stabilize dynamical behaviour of the target system by sensing its state and manipulating the system depending on the sensed-outcomes.
Recently, the real-time stabilization of photonic quantum system \cite{sayrin_real-time_2011} has been also demonstrated by using quantum non-demolition measurements (QNDMs) and the feedback control.
Such an active control of quantum system is a significant step towards realization of advanced quantum information processing \cite{vitali_quantum-state_1998,doherty_quantum_2000,wiseman_quantum_2009}.

In addition to photonic systems, atomic quantum systems are also very attractive from a viewpoint of quantum feedback. 
The measurement-based quantum feedback of an atomic spin ensemble, in particular, enables us to realize important applications such as the quantum memory for continuous-variable system \cite{julsgaard_experimental_2004,sherson_quantum_2006,hammerer_quantum_2010} and the enhancement of the squeezing of the squeezed spin state \cite{trail_strongly_2010} which would improve high precision spectroscopy such as an optical lattice clock \cite{takamoto_optical_2005} or an atomic magneto-meter \cite{budker_optical_2007,huelga_improvement_1997,giovannetti_quantum_2006}.
While there are several experiments that achieve spin-squeezing of an atomic ensemble by using QNDM \cite{Schleier-Smith_states_2010,Appel_mesoscopic_2010, sewell_spin-squeezing_2011, takano_spin_2009, julsgaard_experimental_2001,sherson_deterministic_2006,Chen_conditional_2011}, there has been no report on the measurement-based quantum feedback control in the continuous-variable system.

Here we have achieved unconditional quantum-noise suppression of the atomic quantum fluctuation via successful feedback control of an atomic collective spin state by using QNDMs and coherent manipulations \cite{thomsen_spin_2002}.
Our feedback control successfully converts the conditional quantum-noise reduction into the unconditional one without significant loss of the noise reduction level.
Our result will open the door towards emerging field of quantum information dynamics with feedback \cite{lloyd_quantum-mechanical_1997,leff_maxwells_2002,sagawa_second_2008}.

We note that our measurement-based feedback is not `coherent quantum feedback' \cite{wiseman_all-optical_1994,lloyd_coherent_2000,nelson_experimental_2000,Schleier-Smith_squeezing_2010,leroux_implementation_2010}, with which one controls the dynamics of the target system without measurement device.
Particularly interesting case is the cavity-feedback scheme \cite{Schleier-Smith_squeezing_2010,leroux_implementation_2010}, where an impressive squeezing level for atomic spins is achieved.
Compared with the coherent feedback, our approach does not require any purpose-built interaction.
The power of the quantum measurement enables us to design various time evolution of quantum system.

\begin{figure}[bp]
\includegraphics{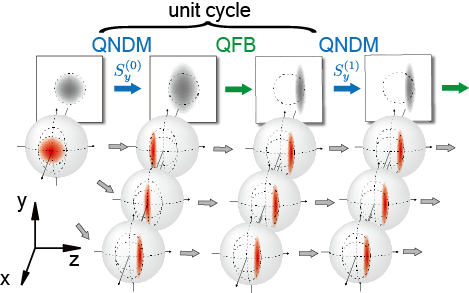}
\caption{(Color online) Schematic illustration of the controlled dynamics.  The atomic spin fluctuation is shown as a circle on the surface of the collective Bloch spheres.  The shadows projected on the screens represent the total spin fluctuations.  The quantum non-demolition measurement (QNDM) brings the initial atomic state into the ensemble of various squeezed spin states, and the quantum feedback (QFB) displaces the squeezed spin states so as to suppress the total spin fluctuation along the $z$-axis.
}
\label{fig:qndm}
\end{figure}

A target quantum system in this work is an atomic collective spin described by a collective spin vector $\bm{J}=\left(J_x,J_y,J_z\right)$.
The uncertainties of spin components, $\delta J_y$ and $\delta J_z$, are jointly limited by the uncertainty relation $\delta J_y\cdot \delta J_z\ge |\braket{J_x}|/2$ due to the commutation relation $[J_y,J_z]=iJ_x$.
In our case, the initial state is given by an $N_A$-partite product state $\ket{\varphi}\equiv\ket{\uparrow_x}\otimes\ket{\uparrow_x}\otimes\cdots\otimes\ket{\uparrow_x}$, where  $\ket{\uparrow_x}$ is an eigenstate of the $x$ component of a single-particle operator of spin-$1/2$.
The state $\ket{\varphi}$ is an eigenstate of $J_x$ with the eigenvalue of $J\equiv N_A/2$, and its standard deviations of the transverse spin components are obtained as $\delta J_y=\delta J_z= \sqrt{J/2}$.
It therefore shows the minimum uncertainty $\delta J_y \cdot \delta J_z=J/2$ equally distributed over any two orthogonal components perpendicular to the $x$ direction, and  $\ket{\varphi}$ is referred to as the coherent spin state.

A QNDM on the atomic collective spin by using a polarized light field as a probe system \cite{kuzmich_atomic_1998,takahashi_quantum_1999} brings the coherent spin state into other quantum states with reduced quantum uncertainty \cite{takano_spin_2009}, which is schematically depicted in Fig.~\ref{fig:qndm}.
The polarization degrees of freedom of light field is described by the photonic Stokes vector $\bm{S}$ whose components are defined as $S_x = (a_y^\dagger a_y-a_x^\dagger a_x)/2$, $S_y = (a_y^\dagger a_x+a_x^\dagger a_y)/2$ and $S_z = (a_y^\dagger a_x - a_x^\dagger a_y )/2i$, where $a_x$ and $a_y$ are annihilation operators of light field linearly polarized along the corresponding directions.
The quantum non-demolition interaction between these two systems is implemented by the Faraday rotation, by which the axis of the polarization of light is rotated by an angle being proportional to the measured atomic collective spin.
The interaction is described by a unitary operator $U_f=\exp\left({-i\chi_f J_z S_z}\right)$,
where $\chi_f$ represents the Faraday rotating angle per unit spin angular momentum.
The QNDM induces a type of inter-particle correlation or entanglement in the collective spin system, and reduces the fluctuation of $J_z$ below $\sqrt{J/2}$.
Such a state is known as a squeezed spin state \cite{kitagawa_squeezed_1993}. 
However, the obtained squeezed spin states is randomly distributed depending on the measurement outcomes over the range of the initial quantum fluctuation of $\ket{\varphi}$ as in Fig.~\ref{fig:qndm}; the total fluctuation along the $z$-axis just after the QNDM is the same as that of the initial coherent state.
The goal of our work in this paper is to suppress this random fluctuation via quantum feedback (QFB) \cite{doherty_quantum_2000,wiseman_quantum_2009}.

\begin{figure}[tbp]
\includegraphics{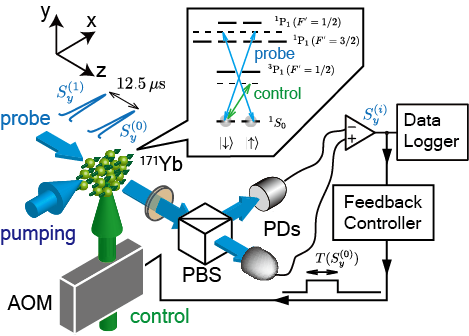}
\caption{(Color online) Apparatus to implement the quantum feedback control with associated energy level structure of $^{171}\text{Yb}$.  AOM, Acoustic-Optical modulator; PBS, Polarization beam splitter; PDs, Photo detectors.  
Horizontally polarized probe pulses illuminate the ensemble of cold ${}^{171}\text{Yb}$ atoms which are spin-polarized along the $x$ direction by an optical pumping beam.
The polarization $S_y^{(0)}$ and $S_y^{(1)}$ of the pulses are measured by a polarimeter which consists of a PBS and two PDs.
A time interval between these two pulses is $12.5\,\mathrm{\mu s}$.
The measurement outcomes are fed into the logger and feedback controller which determines the pulse duration $T(S_y^{(0)})$ of the circularly polarized control pulse.  
}
\label{fig:setup}
\end{figure}

For the experimental implementation of the QFB, we use an ensemble of ytterbium (${}^{171}\text{Yb}$) atoms as the target system.
$\text{Yb}$ has two valence electrons and hence no electronic spin in the ground state.
The $^{171}\text{Yb}$ atoms, in particular, have only a nuclear spin-1/2 in their ground state.
The system therefore constitutes the simplest 2-level energy structure and is robust against decoherence due to residual magnetic field owing to the small nuclear magnetic moment.
There are no effects of an additional rank-2 tensor term which appears in the case of atoms with larger spin, such as usually employed alkaline atoms.
Figure \ref{fig:setup} shows the schematic of experimental setup.
We firstly create the coherent spin state by a $10\text{-}\mathrm{\mu s}$ long optical pumping pulse of circularly polarized light propagating along the $x$ axis and tuned to the ${}^{1}\mathrm{S}_0\leftrightarrow{}^{1}\mathrm{P}_1(F'=1/2)$ transition.
Then we measure the rotation angle of linearly polarized probe pulses $S_y$ which get the information of the atomic collective state via the Faraday rotating interaction.
The probe pulses are tuned to the ${}^{1}\mathrm{S}_0\leftrightarrow{}^{1}\mathrm{P}_1(F'=1/2)$ transition with $\Delta=-2\pi\times160\,\mathrm{MHz}$ detuning, and the mean photon number per the pulse is $N_L=1.3(1)\times 10^6$.
The probe is horizontally polarized before the interaction ($S\equiv |\braket{S_x}|=N_L/2$). 
The Faraday rotating angle with the atoms spin-polarized along the probe direction is $0.16(2)\,\text{rad}$, which corresponds to an effective atom number of $N_A=3.7(4)\times10^5$.

The QFB in this work proceeds in the following way; the output of the polarization detector $S_y^{(0)}$ is fed into both of data logger and feedback controller, then the output is looped back into the atomic system.
The feedback operation is implemented by a fictitious magnetic field \cite{takano_manipulation_2010} which is generated by a circularly polarized `control pulse' propagating along the $y$-axis.
After the QFB, we measure the spin state by applying the second QNDM pulse $S_y^{(1)}$.
A time interval between the first and the second QNDM is $12.5\,\mathrm{\mu s}$ which is sufficiently shorter than the typical decoherence time (several hundred microseconds due to ballistic expansion of cold atomic cloud).

In the conventional normalized quadratures for light and atoms polarized along the $x$-axis $(X_L,P_L)\equiv(S_y,S_z)/\sqrt{S}$ and $(X_A,P_A)\equiv(J_y,J_z)/\sqrt{J}$, we can work with the standard commutation relation $[X_L,P_L]=[X_A,P_A]=i$ and  uncertainty relations $V(X_L)V(P_L)\ge1/4$ and $V(X_A)V(P_A)\ge1/4$ from the commutation relation, where $V(\cdot)$ means the variance.
By using these operators, the Faraday rotating unitary operator $U_f$ can be rewritten as $U_f=e^{-i\chi_f J_zS_z}=e^{-i\kappa P_AP_L}$ with a dimensionless parameter $\kappa \equiv \chi_f\sqrt{SJ}$, which is calculated as \cite{takano_spin_2009}
\begin{align*}
\kappa=\frac{2\sigma_0\Gamma\sqrt{SJ}}{3\pi w_0^2(1+s)}\frac{\Delta}{\Delta^2+(\Gamma/2)^2},\ s=\frac{s_0}{1+(2\Delta/\Gamma)^2},
\end{align*}
where $\Gamma=2\pi\times29\,\mathrm{MHz}$ is the natural linewidth, $\sigma_0=7.6\times10^{-14}\,\mathrm{m^2}$ is the photon-scattering cross section in ${}^{171}\text{Yb}$ atom, the beam waist $w_0=40\,\mathrm{\mu m}$, and the saturation parameter $s_0=7.2$.
In our case, $\kappa$ is estimated as $0.59$.
For these parameters, the damping coefficient \cite{duan_quantum_2000} of light ($\epsilon_L$) and atoms ($\epsilon_A$) can be estimated as $(\epsilon_L,\epsilon_A)=(0.042,0.15)$.

The control pulse, which is tuned to the ${}^{1}\mathrm{S}_0\leftrightarrow{}^{3}\mathrm{P}_1(F'=1/2)$ transition with $-2\pi\times70\,\mathrm{MHz}$ detuning, induces the phase shift and rotates the collective atomic spin vector in the $x$-$z$ plane.
The rotation angle is proportional to both of the time duration and the intensity of the control pulse.
Our feedback controller, which is implemented by a field-programmable gate array, determines the time duration depending on the shot-by-shot measurement outcome.
The intensity of the control pulse is an additional parameter of our feedback, and is used as `feedback gain'.
The negative sign of the gain is assigned for the correct feedback direction to successfully reduce the fluctuation.
The direction of the feedback can be changed by switching the sense of the circular polarization of the control pulse.

\begin{figure}[tbp]
\includegraphics{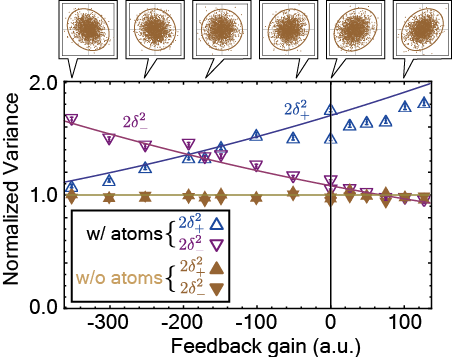}
\caption{(Color online)  
Normalized variances (in light shot noise unit) of the sum ($2\delta_+^2$, $\bigtriangleup$) and the difference ($2\delta_-^2$, $\bigtriangledown$) of two outcomes $S_y^{(0)}$ and $S_y^{(1)}$ as a function of feedback gain (the error bars are the statistical standard deviations $\pm 1\sigma$).  
The value of $1$ of $2\delta_\pm^2$ corresponds to the light shot noise.
The solid curves are the corresponding theoretical predictions from the calculation scaled by a single proportional factor to link the measured intensity of control light pulse to feedback gain.  The insets show measured joint frequency distributions of two sequential QNDM outcomes $S_y^{(0)}$ (horizontal axis) and $S_y^{(1)}$ (vertical axis) with solid curves indicating $3\sigma$ radii. 
}
\label{fig:feedback_corr}
\end{figure}

Figure \ref{fig:feedback_corr} shows the observed correlations between the two sequential QNDM outcomes modified by the feedback control.
We show the normalized variances of the sum and the difference of two outcomes as a function of the feedback gain, which are defined by using $\delta_{\pm}^2\equiv V(X_{L}^{(0)}\pm X_{L}^{(1)})/2=V(S_y^{(0)}\pm S_y^{(1)})/(2S)$.
Each point is calculated based on $10,000$ independent sets of the outcomes with the corresponding feedback gain.
The measured joint frequency distributions of two outcomes $S_y^{(0)}$ (horizontal axis) and $S_y^{(1)}$ (vertical axis) are also plotted at some feedback gains. 
The normalized variances $\delta_\pm^2$ correspond to the width of the distribution along diagonal lines of the individual frame of the insets.
Here we also show the actual data which is obtained without the atoms, and we can see that the normalized variances is close to the expected value.
The excess spin noises are clearly observed for the data with the atoms, regardless of the feedback gain.   
The data with zero feedback gain indicates conditional quantum-noise suppression \cite{takano_spin_2009,takano_manipulation_2010}.
We have confirmed that the observed noise is of quantum origin by several ways \footnote{See Section II in Supplemental Material at [URL] for details}.
A comparison between $\delta_+^2$ and $\delta_-^2$ highlights the effect of the feedback control.
As schematically shown in Fig.~\ref{fig:qndm}, the atomic spin states are expected to be a certain squeezed spin state via the optimal QFB.
It means that the atomic state after the optimal QFB does not depend on the first outcome, the correlation $\delta_+^2 -\delta_-^2$ is therefore expected to be close to zero at the optimal feedback gain $g=g_\text{opt}$. 
In other words, our QFB reduces the quantum fluctuations at the expense of the correlation or the information.
If $|g|<|g_\text{opt}|$ which means that the rotation angle of the mean spin is smaller than the optimal one, the correlation does not reach zero but is still positive. 
On the other hand, if $|g|>|g_\text{opt}|$, the correlation becomes negative one due to the overshooting of the rotation.
Figure \ref{fig:feedback_corr} shows a transition of the correlation from positive to negative with increasing the intensity of control pulse $|g|$ at $g<0$.
This is a direct evidence that the atomic spin states can be manipulated depending on the shot-by-shot measurement outcomes.
The results are in good agreement with our theoretical model \footnote{See Section III in Supplemental Material at [URL] for details of the theoretical model.} (solid lines in Fig. \ref{fig:feedback_corr}).

In order to know the available maximum noise reduction in the unconditional case, we evaluate a conditional parameter. 
From the positive correlation between two QNDM outcomes in the case of no feedback, we define
\begin{align*}
&\xi^2_\text{cond}\equiv \frac{V_\text{cond}-V(S_{y0}^{(1)})}{V(S_{y}^{(1)})-V(S_{y0}^{(1)})},
\end{align*}
where $V_\text{cond}\equiv \underset{g_c}{\text{min}}\ V(S_y^{(1)}+g_c S_y^{(0)})$, and $V(S_{y0}^{(1)})$ corresponds to the variance of the second QNDM in the case of no atoms.
The conditional variance $V_\text{cond}$ means the available minimum variance from the information obtained by the first probe with an ideal readout (i.e., the readout noise is subtracted). 
Therefore, $\xi^2_\text{cond}$ represents how much the atomic spin fluctuations can be suppressed by the correlation between $S_y^{(0)}$ and $S_y^{(1)}$ with ideal QFB.
The optimal value of $g_c$ for the minimization is given by $-\text{cov}(S_y^{(0)},S_y^{(1)})/V(S_y^{(0)})$, where the \text{cov} means the covariance.
The conditional parameter is estimated as $\xi^2_\text{cond}=0.83\pm0.02=-0.80^{+0.11}_{-0.12}\,\text{dB}$, which indicates the successful conditional quantum-noise suppression.
The experimental setup for the conditional quantum-noise suppression is basically the same as that of our previous works \footnote{See \cite{takano_spin_2009,takano_manipulation_2010} and Section II in Supplemental Material at [URL] for further details of our QNDM.}.
We note that in our experiment the coherence is reduced by $0.85$ due to the photon scattering.
At this condition, Kitagawa-Ueda's criterion of the spin squeezing with reduced $J$ is marginally satisfied as $\xi_\text{cond}^2 = 0.83(2) <0.85$.

\begin{figure}[tbp]
\includegraphics{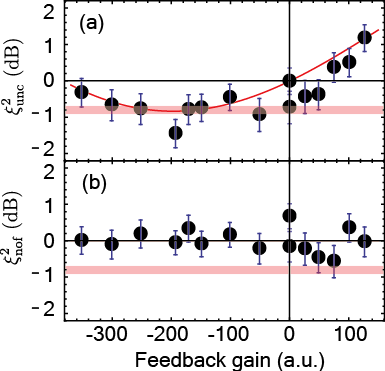}
\caption{(Color online) (a) Unconditional quantum-noise suppression parameter $\xi_\text{unc}^2$ as a function of feedback gain. (b) For comparison, the first QNDM outcomes $S_y^{(0)}$ are used in the calculation of the numerator of the $\xi^2_\text{unc}$, which corresponds to that of no feedback ($\xi_\text{nof}^{2}$).
The error bars represent the statistical error $\pm 1\sigma$. 
The solid curves are the theoretical predictions.
The filled region shows the conditional parameter $\xi^2_\text{cond}=-0.80^{+0.11}_{-0.12}\,\text{dB}$ to indicate the optimal case.
}
\label{fig:feedback}
\end{figure}

With the QFB, a measure of the unconditional quantum-noise suppression is simply the variance of the second outcome $S_y^{(1)}$. 
Here we define the unconditional parameter $\xi^2_\text{unc}$ which characterizes the effect of the QFB as 
\begin{align*}
\xi_\text{unc}^2\equiv \frac{V(S_y^{(1)})-V(S_{y0}^{(1)})}{V(S_y^{(\text{ref})})-V(S_{y0}^{(\text{ref})})},
\end{align*}
where $S_y^{(\text{ref})}$ and $S_{y0}^{(\text{ref})}$ are the outcomes of the polarization measurements with and without atoms in the case of no feedback.
They are measured just before the main sequence \footnote{See Supplementary Fig.~S1c for details}, although it is still possible to use $S_y^{(0)}$ and $S_{y0}^{(0)}$ instead of $S_y^{(\text{ref})}$ and $S_{y0}^{(\text{ref})}$, respectively.
The unconditional parameter $\xi^2_\text{unc}$ represents how much the atomic spin fluctuations are suppressed via the actual QFB with an ideal readout.

Figure \ref{fig:feedback}(a) shows the unconditional parameter as a function of the feedback gain.
While no unconditional quantum-noise suppression is observed for the case of no QFB (b), we clearly observe the maximum reduction of the spin fluctuation in the vicinity of the feedback gain where the correlation $\delta_+^2 -\delta_-^2$ is close to zero (see also Fig.~\ref{fig:feedback_corr}), as expected.
This demonstrates the unconditional quantum-noise suppression as a result of the successful QFB.
In addition, we have successfully observed the features such as maximum (and near-optimal) suppression at some negative gain, no suppression at zero and too much negative gain, and also the increase of the noise at positive gain.
The overall features of the observation in Fig.~\ref{fig:feedback} are in good agreement with the theoretical model \footnote{See Section III in Supplemental Material at [URL] for details of the theoretical model.}.

The important feature of the QFB is that we can manipulate the dynamics of the target quantum system via consecutive feedback.
While the achieved quantum-noise suppression and the successful feedback steps in this study are limited because of the absorption of photons in probe pulses and the ballistic expansion of the atomic cloud, these problems are not of the fundamental origin and can be overcome by realistic improvement such as using an optical trap system to increase the optical density of atomic cloud.\footnote{See Section III-C in Supplemental Material at [URL] for details of the theoretical model.}.
We believe that our results pave the way for more advanced QFB \cite{doherty_quantum_2000,thomsen_spin_2002,wiseman_quantum_2009} of the collective spin system.

In summary, we have experimentally demonstrated the unconditional quantum-noise suppression via quantum feedback control in an unconditional manner.
We note that our feedback controller can be regarded as a quantum analog of Maxwell's demon \cite{lloyd_quantum-mechanical_1997,leff_maxwells_2002,sagawa_second_2008,toyabe_experimental_2010}.
While the original classical demon only suppresses thermal fluctuations, our demon can suppress quantum, thermal, and technical noise equally if the measurements give him the information of the corresponding noise.
Since the effect of thermal and technical noise is negligibly small \footnote{See Section `Performance of QNDM' for details.} in the experiment, we have confirmed that the observed noise reduction here is of quantum origin.

\begin{acknowledgements}
We acknowledge S.~Uetake and K.~Takeda for technical assistance and T.~Takano for valuable comments.  
This work is supported by the Grant-in-Aid for Scientific Research of JSPS (No.~18204035, 21102005C01 (Quantum Cybernetics)), GCOE Program ``The Next Generation of Physics, Spun from Universality and Emergence'' from MEXT of Japan, and World-Leading Innovative R\&D on Science and Technology (FIRST).
\end{acknowledgements}


%

\appendix

\section{Experimental setup}

\begin{figure*}[tbp]
\includegraphics{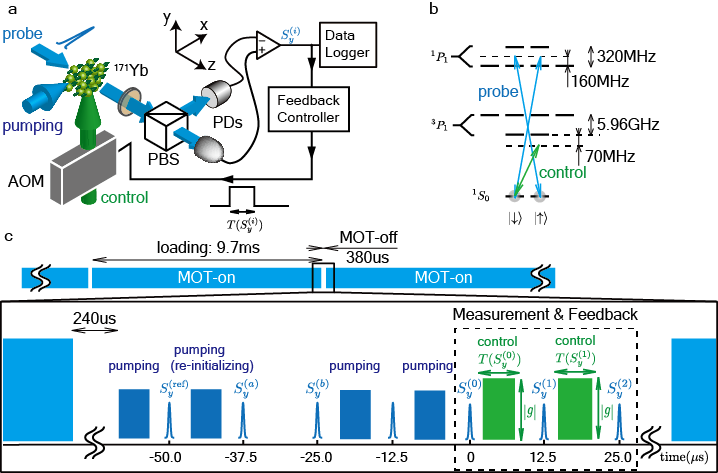}
\caption{{\bf Schematic illustration of the experimental setup.} 
 (a) Experimental apparatus. (b) Energy level structure of ${}^{171}\text{Yb}$ and associated laser frequencies.  (c) Main-sequence of the experiment.}
 \label{fig:setup_SM}
\end{figure*}

In this study, we use a cold ${}^{171}\text{Yb}$ atomic cloud as the target system of our quantum feedback control.
The whole setup and timing sequence are shown in Fig.~\ref{fig:setup}.
The measurements are performed within the ``MOT-off'' periods, where the magnetic and light fields responsible for cooling and trapping are switched off.
During the period, we firstly create the coherent spin state by a $10\text{-}\mathrm{\mu s}$ long optical pumping pulse of circularly polarized light propagating along the $x$ axis and tuned to the ${}^{1}\mathrm{S}_0\leftrightarrow{}^{1}\mathrm{P}_1(F'=1/2)$ transition.
Then we measure the rotation angle of linearly polarized probe pulses which get the information of the atomic collective state via the Faraday rotating interaction.
The probe pulses are tuned to the ${}^{1}\mathrm{S}_0\leftrightarrow{}^{1}\mathrm{P}_1(F'=1/2)$ transition with $-2\pi\times160\,\mathrm{MHz}$ detuning as shown in Fig.~S1(b).

Figure S1(c) shows the main-sequence of the experiment.
Here $S_y^{(\cdot)}$ denotes the outcome which is obtained within the main-sequence.
Within the `Measurement \& Feedback' region depicted by the dashed square in Fig.~S1(c), we perform quantum feedback experiment by utilizing the control pulse, which is circularly polarized and tuned to the ${}^{1}\mathrm{S}_0\leftrightarrow{}^{3}\mathrm{P}_1(F'=1/2)$ transition with $-2\pi\times70\,\mathrm{MHz}$ detuning.
The time-width $T$ of the control pulse is determined by the foregoing outcome $S_y^{(i)}$, and the function is implemented by the field-programmable gate array as the `Feedback Controller' and by the `AOM' (Acoustic-Optical Modulator) as schematically shown in Fig.~S1(a).
In this system, we can control the feedback gain $g$ by changing the intensity and the polarization of the control pulse.
After the measurement and control period, the new atomic cloud is formed for the next sequence at a rate of about $100\,\mathrm{Hz}$.
The main sequence was repeated $10,000$ times with a certain value of the feedback gain.
This is a unit experimental run of the experiment, and is also repeated with various values of the feedback gain.

We also measure atom-free data by using the same setup to estimate the amount of light noise in the polarimeter.
This light noise is subtracted from the full observed noise in calculating the noise suppression $\xi^2$.

\section{Performance of QNDM}

\begin{table*}[tbp]
\begin{center}
\begin{tabular}{|c|c|c|c|}
\hline
\makebox[20mm][c] &\makebox[40mm][c]{(i) without atoms}& \makebox[40mm][c]{(ii) independent spin states}& \makebox[40mm][c]{(iii) identical spin states}\\
\hline\hline
$V(S_y^{(0)})$ & $3.21(1)\times10^5$ & $4.43(2)\times10^5$ & $4.44(2)\times10^5$ \\
\hline
$V(S_y^{(1)})$ & $3.22(1)\times10^5$ & $4.44(2)\times10^5$ & $4.43(2)\times10^5$ \\
\hline
$\Delta_+^2$ & $3.29(1)\times10^5$ & $4.73(2)\times10^5$ & $5.39(2)\times10^5$ \\
\hline
$\Delta_-^2$ & $3.15(1)\times10^5$ & $4.15(1)\times10^5$ & $3.49(1)\times10^5$ \\
\hline
$\xi^2_\text{cond}$&-&0.98(2)&0.83(2)\\
\hline
\end{tabular}
\end{center}
\caption{{\bf Variances and conditional noise-suppression parameters corresponding to Fig.~\ref{fig:setup_SM} (i)-(iii).}  All values are calculated from 150,000 polarimeter outcomes, which are converted to the corresponding photon number differences.  The outcomes are collected simultaneously with the quantum feedback experiments.  The numbers in parentheses denote $1\sigma$ statistical errors.  
}
\end{table*}

\begin{figure}[bp]
\includegraphics{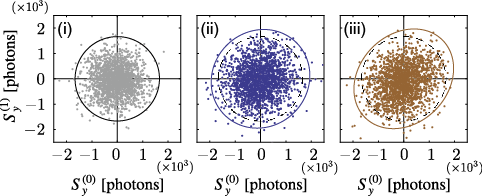}
\caption{{\bf Joint distributions of the two sequential QNDM-outcomes $S_y^{(0)}$ and $S_y^{(1)}$ without any feedback control.}  The measurement is performed with (i) no atoms, (ii) independent coherent spin states prepared by an additional pumping pulse between two probe pulses, and (iii) identical collective spin states.  Solid curves indicate $3\sigma$ radii and dashed curves in (ii) and (iii) are the same as the solid curve in (i).
}
\end{figure}

Figure~\ref{fig:setup_SM}(i)-(iii) show the correlation plots of the two sequential quantum non-demolition measurement (QNDM) outcomes with (i) no atoms, (ii) independent coherent spin states, and (iii) identical collective spin states in the case of no feedback .
In our experiments, the pairs of outcomes $S_y^{(0)}$ and $S_y^{(1)}$ corresponding to the case (ii) and (iii) are obtained simultaneously with that for quantum feedback experiments as shown in Fig.~\ref{fig:setup_SM}(c); the pair of $S_y^{(\text{ref})}$ and $S_y^{(a)}$ corresponds to the case (ii), and the pair of $S_y^{(a)}$ and $S_y^{(b)}$ corresponds to the case (iii).
Here we confirm that the observed variances $V(S_y^{(0)})$ and $V(S_y^{(1)})$ are consistent with the quantum noise limited values.
Each probe pulse of the QNDMs has on average $N_L= 1.3(1)\times 10^6$ photons.
From this value, we can estimate the expected variance in the case of no atoms (i) as $V(S_y) = N_L/4= 3.3(3)\times10^5 $, which is consistent with the observations in Table I (i).
The atom number in the probe region is determined from the measurement of the Faraday rotation angle $\theta$ with the sample spin-polarized along the probe direction. 
The Faraday rotating angle per unit spin angular momentum $\chi_f$ links the effective number of atoms with the Faraday rotating angle as $\theta=\chi_f N_A/4$.
A typical value of $\theta = 0.16(2)$ rad for the rotation angle corresponds to the effective atom number of $3.7(4)\times 10^5$. 
We can estimate the expected quantum-noise-limited variance of the polarimeter outcomes with the sample spin-polarized perpendicular to the propagation axis of the probe pulse as $V(S_y) = N_L/4+(\chi_f \cdot N_L/2)^2\cdot N_A/4= 4.4(4)\times10^5 $, which is consistent with the experimentally obtained variances $V(S_y^{(0)})$ and $V(S_y^{(1)})=4.4\times10^5$ listed in Table I(ii) and (iii).

Here the correlation between two QNDM outcomes $S_y^{(0)}$ and $S_y^{(1)}$ is clearly seen in Fig.~6(iii).
Quantum noise reduction is now only possible in a conditional way via the correlation.
The standard deviation $\Delta_\pm$ which is defined as $\Delta_\pm^2\equiv V(S_y^{(0)}\pm S_y^{(1)})/2$ characterizes the correlation, and corresponds to the width of the distribution along diagonal lines of the individual frame of Fig.~\ref{fig:setup_SM}.
The obtained $\Delta_\pm^2$ is summarized in Table I with statistical errors.
One can clearly see a reduction of $\Delta_{-}$ as well as an enhancement of $\Delta_{+}$ in the case of identical spin state (iii), compared with $\Delta_\pm$ of independent spin state (ii) and also the respective $V(S_y^{(0)})$ and $V(S_y^{(1)})$.

In order to check that the observed noise is of quantum origin, we also evaluate the conditional quantum-noise suppression with applying an additional pumping pulse during the two probe pulses $S_y^{(0)}$ and $S_y^{(1)}$.
The pumping pulse resets the collective atomic spin state and thus the correlation due to the quantum fluctuation is expected to disappear as a result of the re-initialization. 
Consequently, the correlation, if observed even with the application of the pumping pulse, should indicate the contribution from the classical noise.
In our experiment, we minimize the possible residual correlation by adjusting an incident angle of the pumping pulse, which results in the conditional parameter of  $\xi^2_\text{cond}=0.98\pm0.02=-0.068\pm0.10$ dB for the independent coherent spin state shown in Fig.~\ref{fig:setup_SM}(ii).
This result confirms that the effect of thermal and technical noise is too small to mask the quantum fluctuations and our QNDM has sufficient sensitivity to access the quantum fluctuations.
Possible origin of the technical noise is the residual magnetic field and the probe pointing fluctuation.
The slight reduction of the conditional noise-suppression level compared with our previous work \cite{takano_spin_2009} comes from the slight decrease of the atom number, and has nothing to do with the performance of the measurement-based quantum feedback control.
We note that an increase of the conjugate quadrature noise due to the QNDM was observed in our previous work using essentially the same experimental setup \cite{takano_manipulation_2010}.

\section{Theoretical model}

\subsection{QND interaction with scattering loss}

After the probe pulse passes through the atomic ensemble, the QND interaction under the influence of the scattering loss changes the canonical operators $X_L$ and $P_A$ \cite{duan_quantum_2000}:
\begin{align}\label{original1}
X'_L&=\sqrt{1-\epsilon_L}(X_L+\kappa P_A)+\sqrt{\epsilon_L} E_L,\\
P'_A&=\sqrt{1-\epsilon_A}P_A + \sqrt{\epsilon_A} E_A,\nonumber
\end{align}
where the symbols with (without) a prime denote the operators after (before) the interaction, $E_L$ and $E_A$ are the standard vacuum operators with variance $1/2$, and $\epsilon_L$ and $\epsilon_A$ are the damping coefficients.
The vacuum operators represent the contributions from unpolarized components due to the scattering.
There are two essential assumptions to derive Eq.(\ref{original1}): (i) one-dimensional model is appropriate, in other words, the interaction volume is of a pencil shape with Fresnel number $F\sim 1$  and (ii) the excited atoms decay into either of the two spin states in the ground state.
The scattered photons are emitted into the same spatial mode with the probe pulse because of the former assumption, and thus are fed into the polarimeter. 
This effect is represented by the contribution of $E_L$.
Similarly, all atoms which absorb a photon in the probe pulse come back to either of the two spin states in the ground state because of the latter assumption, therefore the atoms are not lost but partially depolarized due to the photon absorption-and-scattering process.
The effect is represented by the contribution of $E_A$.

In our experiment, the Fresnel number $F\sim 10$ is much larger than unity and there are no other ground states; i.e., the latter (ii) is appropriate, but the former (i) is not.
The QND interaction in our experiment is therefore described by the following transformation:
\begin{align}\label{eq:trans}
X'_L&=\sqrt{1-\epsilon_L}(X_L+\kappa P_A),\\
P'_A&=\sqrt{1-\epsilon_A}P_A + \sqrt{\epsilon_A} E_A.\nonumber
\end{align}

\subsection{Uncorrelated noise}

Variance of the sum ($\delta_+^2\equiv V(X_L^{(0)}+X_L^{(1)})/2$) and the difference ($\delta_-^2\equiv V(X_L^{(0)}-X_L^{(1)})/2$) characterize the correlation between the two sequential outcomes.
For the ideal case ($\epsilon_A=\epsilon_L=0$) without feedback ($g=0$), the difference $X_L^{(0)}-X_L^{(1)}$ becomes $X_{L0}^{(0)}-X_{L0}^{(1)}$ and this is equal to the case without atoms ($\kappa=0$).
The variance $\delta_-^2$ without feedback ($g=0$) can also be calculated by using the Eq.(\ref{eq:trans}) as
\begin{align*}
\delta_-^2(\kappa,\epsilon_A,\epsilon_L)=\frac{1-\epsilon_L}{2}\left(1+\kappa^2(1-\sqrt{1-\epsilon_A})\right),
\end{align*}
and $\delta_-^2(\kappa,\epsilon_A,\epsilon_L)\lesssim V(X_{L0}^{(i)})=1/2$ holds over a wide range of the parameters around $(\kappa,\epsilon_A,\epsilon_L)=(0.59,0.15,0.042)$, e.g., $\delta_-^2(0.59,0.15,0.042)=0.49$.
However, as shown in Table I (i) and (iii), the experimentally observed $\Delta_-^2 = 3.49(1)\times10^5$ is slightly larger than the variance of the light shot noise $3.21(1)\times10^5$ which is measured without atoms.
The results suggest that the measurement outcomes include some additional uncorrelated noise, which cannot be cancelled out by taking the difference between two sequential outcomes similarly to the light shot noise.
The amount of the additional noise can be estimated as $V(\tilde{E}_L)=(3.49(1)\times10^5-3.21(1)\times10^5)/2=0.14(1)\times10^5$.
We emphasize that the kind of noise cannot be reduced by using the feedback in any way, and it just decreases the amount of the noise reduction.
We can effectively take into account the presence of uncorrelated noise by adding the uncorrelated noise terms in Eq.(\ref{original1}) as follows:
\begin{align}\label{eq:revised}
X'_L&=\sqrt{1-\epsilon_L}(X_L+\kappa P_A)+\sqrt{\epsilon'_L} E_L,\\
P'_A&=\sqrt{1-\epsilon_A}P_A + \sqrt{\epsilon_A} E_A.\nonumber
\end{align}
where a pair of $\epsilon'_L$ and $E_L$ (with variance $1/2$) represents the contribution form the additional uncorrelated noise.
The term $\sqrt{\epsilon'_L}E_L$ includes not only the scattering photons as the standard vacuum operator in Eq.(\ref{original1}) or Ref.\cite{duan_quantum_2000}, but also the contributions from all uncorrelated noise in the presence of atoms.
We can estimate $\epsilon'_L=0.098$ from the modified model Eq.(\ref{eq:revised}) with the experimental results $\Delta_-^2 = 3.49(1)\times10^5$ and the variance of the light shot noise $3.21(1)\times10^5$.

\subsection{Quantum feedback}

\begin{figure}[b]
\includegraphics{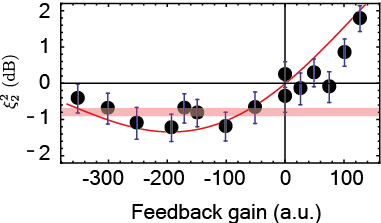}
\caption{{\bf Two-cycle feedback results as a function of feedback gain.}  The error bars represent the statistical error $\pm 1\sigma$ and the solid curve is the theoretical predictions.
The filled region shows the conditional quantum-noise suppression $\xi^2_\text{cond}=-0.80^{+0.11}_{-0.12}\,\text{dB}$ to indicate the optimal squeezing with 1-cycle feedback (Fig.~4(a)).
}
\label{fig:multiFB}
\end{figure}

Adding the feedback term to Eq.(\ref{eq:revised}) yields the following formula:
\begin{align}\label{eq:feedback}
X'_L&=\sqrt{1-\epsilon_L}(X_L+\kappa P_A)+\sqrt{\epsilon'_L} E_L,\\
P'_A&=\sqrt{1-\epsilon_A}P_A+\sqrt{\epsilon_A}E_A\nonumber\\
&\qquad\qquad +g\cdot(1-\epsilon_A)(X'_L-x_0),\nonumber
\end{align}
where the third term of $P'_A$ represents the effect of feedback.
The pre-arranged target value $x_0$ has set to be about $3\cdot\sqrt{V(X'_L)}$ in the experiment.
The absorption makes effective feedback gain to be $1-\epsilon_A$ times smaller than that in the case of no scattering loss.

We have also done the multi-cycle feedback in the experiment, as shown in the timing sequence (Fig.~\ref{fig:setup_SM}(c)).
We can express the multi-cycle feedback by using the following recurrence formula which corresponds to the generalized form of Eq.(\ref{eq:feedback}) as
\begin{align}\label{eq:multiFB}
X_L^{(i)}&=\sqrt{1-\epsilon_L}(X_{L0}^{(i)}+\kappa P_A^{(i)}) + \sqrt{\epsilon'_L} E_L^{(i)},\\
P_A^{(i)}&=\sqrt{1-\epsilon_A}P_A^{(i-1)}+\sqrt{\epsilon_A}E_A^{(i)}\nonumber\\
&\qquad +g\cdot(1-\epsilon_A)(X_L^{(i-1)}-x_0)\cdot F[X_L^{(i-1)},x_0],\nonumber\\
&\text{where } F[a,b]=\begin{cases}1&a\le b\\0&a>b\end{cases}.\nonumber
\end{align}
Here the variances are normalized as $V(X_{L0}^{(i)})=V(P_A^{(0)})=V(E_L^{(i)})=V(E_A^{(i)})=1/2$, and the working range of our feedback is limited as represented by the function $F[X_L^{(i-1)},x_0]$.
We define the conditional noise suppression parameter with the $i$-cycle feedback as $\xi^2_i\equiv(V(X_L^{(i)})-1/2)/(V(X_{L}^{(0)})-1/2)$.

Figure~\ref{fig:multiFB} shows the experimental result of $\xi_2^2$, and it is again in good agreement with the theoretical prediction calculated by Eq.(\ref{eq:multiFB}).
Note that the multi-cycle feedback is performed with the pre-fixed feedback gain, which is known as not the optimal one \cite{thomsen_spin_2002}.
Multi-cycle feedback is expected to improve the noise suppression and broaden the gain dependence.

In this study, the achieved quantum-noise suppression and the successful feedback steps are limited because of the absorption of photons in probe pulses and the ballistic expansion of the atomic cloud.
The realistic way to overcome the problems is to use an optical trap system.

We can characterize the optical response of our collective spin system by using a complex valued susceptibility with taking into account of transition probabilities of associated transitions of ${}^{171}\text{Yb}$.
By using the model, the coupling coefficient $\kappa$ and the atomic damping coefficient $\epsilon_A$ can be represented as follows:
\begin{align*}
\kappa&=\frac{2\sigma_0\Gamma\sqrt{SJ}}{3\pi w_0^2(1+s)}\frac{\Delta}{\Delta^2+(\Gamma/2)^2},\\
\epsilon_A&\simeq\frac{\sigma_0\Gamma S}{\pi w_0^2(1+s)}\frac{\Gamma/2}{\Delta^2+(\Gamma/2)^2},\\
s&=\frac{s_0}{1+(2\Delta/\Gamma)^2}.
\end{align*}
The ratio between coupling strength and the atomic damping coefficient is therefore given by
\begin{align*}
\kappa/\epsilon_A\simeq \frac{2}{3}\cdot \sqrt{\frac{J}{S}}\cdot \frac{\Delta}{\Gamma/2}.
\end{align*}
This result means that we should increase $\sqrt{J/S}$ with some fixed value of $\kappa\propto\sqrt{JS}$ in order to improve the feedback system. 
One of the most promising way to increase the effective $J$ is to increase the optical density by using the optical trap system.

\subsection{Feedback control with partially polarized spins}

In the theoretical model described above, we assume that the unpolarized component of atoms ($\sqrt{\epsilon_A} E_A$) is not changed by the feedback control.
Here we consider the reliability of the assumption.
First of all, it is likely that we cannot get any information of the unpolarized spins from the first outcome $X_L^{(0)}$, and thus the feedback control, which depends on $X_L^{(0)}$, only applies some random spin-rotation and does not contribute the noise reduction of the unpolarized spins.
In addition, even if we could get the information of the unpolarized spins from $X_L^{(0)}$, the amount of the information is much smaller than the light shot noise because the upper bound of the signal to noise (signal from the unpolarized spins to the light shot noise) ratio can be estimated as $\kappa^2\cdot \epsilon_A\simeq 0.05$.

\begin{figure*}[htb]
\includegraphics{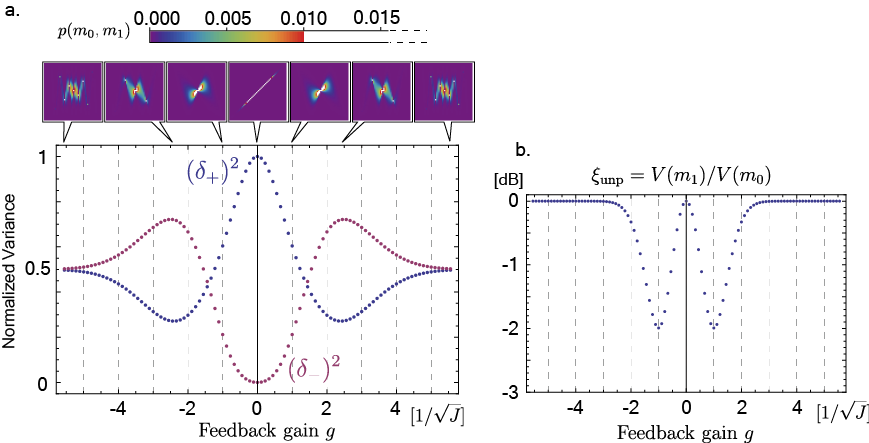}
\caption{{\bf Numerical calculation of the measurement-based feedback for completely mixed states.} 
$m_0$ and $m_1$ correspond to $S_y^{(0)}$ and $S_y^{(1)}$ in the experiment (in the case that the read-out noise can be negligible), respectively. (a) Variances of the sum ($\delta_+^2$) and the difference ($\delta_-^2$) calculated with $\rho_{100}$ as a function of feedback gain, and corresponding joint probability distributions (insets).  (b) Unconditional noise reduction after the feedback for $\rho_{100}$ as a function of the feedback gain, which is essentially the same as the plot in Fig.~4(a).}
\label{fig:corr}
\end{figure*}

If we could obtain the information of the unpolarized spins with sufficient signal to noise ratio, the noise of the unpolarized spins can be reduced by the feedback.
However, the gain dependence with the unpolarized spins is supposed to be drastically different from that with polarized spins.
For the unpolarized spins, the positive and negative feedback equally reduce the spin fluctuation whereas the negativity of the feedback gain is essential for the polarized spins.
This is because there are no special directions for the unpolarized spins except for the measurement axis, whereas the mean spin direction is well-defined for the polarized spins.
In addition, the amount of the noise reduction for the unpolarized spins is negligibly small comparing with that for the polarized spins in the parameter range of the experiment.
This can be qualitatively understood in the following way.
In order to change the distribution of the $z$-component of angular momentum $m_z$, we displace $m_z$ by rotating the spins in our setup.
The amount of the displacement with small rotation angle $\beta$ can be estimated as $\delta m_z =j\beta$, where $j$ is the total angular momentum. 
The polarized spin-$1/2$ system with $N$ atoms can be represented by the number of $(N+1)$ state-vectors with $j=N/2\equiv J$ while the unpolarized mixed state requires $2^N$ state-vectors whose total angular momentum is mostly smaller than $J$.
For the noise reduction with the unpolarized spins, the rotation angle $\beta$ therefore has to be larger than that with the polarized spins to achieve the same amount of the displacement.
The mean total angular momentum of the unpolarized mixed state can be estimated as $\sqrt{\braket{J^2}}\sim\mathcal{O}(\sqrt{\braket{J_x^2}+\braket{J_y^2}+\braket{J_z^2}})\sim\mathcal{O}(\sqrt{J})$.
In our feedback control, the rotation angle $\beta$ is given by the measurement outcome $m_z$ and the feedback gain $g$ as $\beta=gm_z$.
For the polarized spins, the feedback with the optimal gain $g_o$, which gives the maximum noise suppression, is determined by $\beta=g_o m_z$ and $J\beta=-m_z$, then the optimal feedback gain is roughly $|g_o|\sim\mathcal{O}(1/J)$.
This is consistent with the experimental parameter in our feedback control.
On the other hand the optimal gain for the unpolarized spin state, say $g_o'$, is much larger because the larger rotation angle is required with the same outcome $m_z$.
Similarly, the optimal value is determined by the inverse of the mean total angular momentum as $|g'_o|\sim\mathcal{O}(1/\sqrt{J})$.
Since $J\sim10^5$ in the experiment, the value of $|g'_o|$ estimated from this argument is more than $100$ times larger than the value of $|g_o|$.

The asymmetric behaviour of polarized spins is clearly observed in Fig.~3 and 4.
It confirms that the observed spin noise reduction arises from the feedback control of the polarized component, and that the most of the unpolarized component is not changed within the parameter range of the feedback gain, which is around the optimal one for polarized spins $|g_o|\ll |g'_o|$.

In the following part of this section, we analyze the effect of the feedback control on the unpolarized spins more quantitatively.
In order to see the effect of noise suppression on the unpolarized spins, let us define the complete basis labelled by $j$, $m_z$, and an additional quantum number $d$.
Using the basis $\{\ket{j,m_z,d}\}$, we can express the unpolarized $N$-spin state as
\begin{align}
\rho_N&\equiv I_{2^N}/2^N\\
& =\frac{1}{2^N}\sum_{j=j_0}^{N/2}\sum_{m_z=-j}^{j}\sum_{d=1}^{P_{Nj}}\ket{j,m_z,d}\!\!\bra{j,m_z,d},\nonumber
\end{align}
where $j_0=0$ with the even $N$, and $j_0=1/2$ with the odd $N$.
The magnitude $j$ is either an integer or half-integer and is less than or equal to $J=N/2$.
The quantum number $d=1,\cdots, P_{Nj}$ distinguishes between the $P_{Nj}$-degenerated subspaces \cite{mikhailov_addition_1977}:
\begin{align}
P_{Nj}=\frac{N!(2j+1)}{(N/2-j)!(N/2+j+1)!}.
\end{align}
The mixed state $\rho_N$ is already given by the ensemble of the quantum states which are characterized by $m_z$.
Since $m_z$ corresponds to the first measurement outcomes $S_y^{(0)}$ or $X_L^{(0)}$, we use the character $m_0$ instead of $m_z$ for the sake of convenience.
Note that we ignore the readout noise here, but the approximation just emphasizes the gain-dependence for the unpolarized spins.
We can express the state after the feedback as
\begin{align*}
&\rho'=\\
&\frac{1}{2^N}\sum_{j=j_0}^{N/2}\sum_{m_0=-j}^{j}\sum_{d=1}^{P_{Nj}}U(g m_0)\ket{j,m_0,d}\!\!\bra{j,m_0,d}U^\dagger(g m_0),
\end{align*}
where $g$ is the feedback gain, and $U(\beta)\equiv\exp\left(-iJ_y\beta\right)$ represents the rotation about the $y$-axis with the rotation angle $\beta$.
Our interest is in the probability of getting the outcome $m_1$, which corresponds to $S_y^{(1)}$ or $X_L^{(1)}$, by measuring the observable $J_z$ of the state $\rho'$.
Corresponding measurement operator is given by $M_{m_1}=\sum_{j=j_0}^{N/2}\sum_{d=1}^{P_{Nj}}\ket{j,m_1,d}\!\!\bra{j,m_1,d}$.
The joint probability distribution of $m_1$ (after the feedback) and $m_0$ (before the feedback) can be calculated with Wigner D function \cite{sakurai_modern_1994} $D_{m_0,m_1}^{(j)}(\beta)=\braket{j,m_1,d| e^{iJ_y\beta}|j,m_0,d}$ as
\begin{align}
p(m_0,m_1)&\equiv\text{Tr}[M_{m_1}^\dagger M_{m_1}\rho']=\text{Tr}[M_{m_1}\rho']\\
&=\frac{1}{2^N}\sum_{j=j_0}^{N/2} P_{Nj}|D_{m0,m_1}^{(j)}(g m_0)|^2.\nonumber
\end{align}
Probability distributions $p(m_0)$ and $p(m_1)$ can also be calculated as $p(m_1)=\sum_{m_0}p(m_0,m_1)$ and $p(m_0)=\sum_{m_1}p(m_0,m_1)$.

We can estimate various variances $V(\cdot)$ by using the probability distributions.
Figure \ref{fig:corr}(a) shows the numerically estimated variances of the sum ($\delta_+^2\equiv V(m_0+m_1)/2J$) and the difference ($\delta_-^2=V(m_0-m_1)/2J$) as a function of the feedback gain $g$ for the completely mixed state $\rho_{N}$ with $N=100$.
One can clearly see the variances change symmetrically about the feedback gain.
Figure \ref{fig:corr}(b) also shows the noise reduction due to the feedback for $\rho_{100}$.
Here we define the noise reduction parameter of the unpolarized spins as $\xi_\text{unp}=V(m_1)/V(m_0)$, which corresponds to the unconditional squeezing parameter of the polarized spins.
It is suggested that the optimal gain $g'_o$ is $\pm1/\sqrt{J}$ in single feedback for $\rho_{100}$, and the result is consistent with the above estimation of $J\sim \mathcal{O}(1/\sqrt{J})$.

\end{document}